# Stealthy Peers: Understanding Security Risks of WebRTC-Based Peer-Assisted Video Streaming

Siyuan Tang
*Indiana University Bloomington*
tangsi@iu.edu

Eihal Alowaisheq
*King Saud University*
ealowaisheq@ksu.edu.sa

Xianghang Mi[§]
*University of Science and Technology of China*
xmi@ustc.edu.cn

Yi Chen
*Indiana University Bloomington*
chen481@iu.edu

XiaoFeng Wang
*Indiana University Bloomington*
xw7@indiana.edu

Yanzhi Dou
*Independent researcher*
aaron.yzdou@gmail.com

*Abstract*—As an emerging service for in-browser content delivery, peer-assisted delivery network (PDN) is reported to offload up to 95% of bandwidth consumption for video streaming, significantly reducing the cost incurred by traditional CDN services. With such benefits, PDN services significantly impact today's video streaming and content delivery model. However, their security implications have never been investigated. In this paper, we report the first effort to address this issue, which is made possible by a suite of methodologies, e.g., an automatic pipeline to discover PDN services and their customers, and a PDN analysis framework to test the potential security and privacy risks of these services. Our study has led to the discovery of 3 representative PDN providers, along with 134 websites and 38 mobile apps as their customers. Most of these PDN customers are prominent video streaming services with millions of monthly visits or app downloads (from Google Play). Also found in our study are another 9 top video/live streaming websites with each equipped with a proprietary PDN solution. Most importantly, our analysis on these PDN services has brought to light a series of security risks, which have never been reported before, including free riding of the public PDN services, video segment pollution, exposure of video viewers' IPs to other peers, and resource squatting. All such risks have been studied through controlled experiments and measurements, under the guidance of our institution's IRB. We have responsibly disclosed these security risks to relevant PDN providers, who have acknowledged our findings, and also discussed the avenues to mitigate these risks.

## 1. Introduction

With the ever-expanding footprint of video streaming in Internet traffic (projected to reach 82% in 2022 [17]), the techniques and infrastructures for effective and efficient delivery of video content become increasingly important. Past decades have witnessed the incessant growth of content delivery networks (CDNs) for distributing and caching web content across different geolocations, which however is considered to be rather expensive for video streaming. Further, CDNs today have been constrained by their deployment that may not be adequate for serving video-on-demand (VOD) or live-streaming users around the world, given that even the largest CDN provider has only 325K servers located in 1.4K networks by April 2021 [18]. An answer to these challenges is the emergence of *Peer-assisted Delivery Network* (*PDN*) that utilizes a Peer-to-Peer (P2P) protocol (i.e., *WebRTC* [14]) to facilitate video transmission among web browsers. This alternative is considered to be more scalable and cost-effective: for example, Peer5, one of the most popular PDN services, claims to be able to offload 95% bandwidth cost for its customers [30]. PDNs can be easily integrated into today's video streaming infrastructures: a video streaming website simply needs to subscribe to a PDN service and embed the respective PDN Javascript SDK into its video streaming web pages or apps. Then, an ad hoc P2P network among their viewers will be built up, with all coordination and management tasks handled by the PDN provider behind the scene. On the other hand, given known weaknesses of other P2P networks [39], [48], [58], [74], this new content delivery model may have serious security and privacy implications, which however have never been fully understood. Specifically, one may ask whether user consent are clearly and freely communicated when involving video viewers into a PDN network, whether and to what extent a PDN peer's video streaming activities will be exposed to unknown or even malicious parties, whether serving as a PDN peer will incur non-negligible computing and bandwidth overhead, and whether video segments relayed by untrustworthy PDN peers have been properly protected.

**Challenges and solutions**. Answering these questions requires an in-depth analysis of existing PDN services, which turns out to have multiple challenges. First, PDN providers tend to hide their technical mechanisms with few or no publicly available technical documents as well as heavily obfuscated client-side PDN libraries, making it challenging to understand and evaluate existing PDN systems. Second,

PDN services take a hybrid model of combing normal CDN traffic with P2P traffic, and most PDN activities are mixed with heterogeneous in-browser web activities, rendering them stealthy and hard to detect. Making it more complicated is that PDN services are dynamically loaded when visiting a video website or app, and a PDN customer may set various preconditions before loading the PDN services, e.g., the PDN traffic of Douyu TV (a live streaming platform) is only observable through IP addresses located in China.

Despite these challenges, we performed the first systematic study on PDN's security implications. Our study started by collecting publicly available PDN providers and their customers (e.g., video streaming websites). More specifically, our study focuses on 3 representative PDN providers which were found to be most popular in terms of their daily DNS query volume as learned from passive DNS [21]. Then we identified signatures for fingerprinting PDN SDKs (JavaScript and Android) and moved to build up a signature-based PDN customer detector. Our detector leads to the discovery of 134 websites and 38 Android apps, along with another 9 popular video streaming websites integrated with proprietary PDN solutions (*private PDN services*). Regarding the popularity of PDN customers, 92 (69%) websites were found to have over 1 million monthly visits, e.g., RT News (*rt.com*) and Clarin (*clarin.com*), while 25 (66%) apps have over 1 million downloads on Google Play, e.g., ZEE5 TV (*com.graymatrix.did*) and iFlix (*iflix.play*). Upon those PDN services and customers, we conducted a comprehensive analysis in an attempt to identify fundamental security risks and privacy concerns. This has been made possible by a PDN analyzer we built up to automatically run predefined security tests given a PDN service.

**Security discoveries**. Our study on PDN services has brought to light significant security implications of these services, which have never been reported before. Particularly, we found that public PDN services are seriously vulnerable, due to not only misconfiguration on the side of the PDN customer, but also insufficient protection by the PDN provider. More specifically, all public PDN services we discovered are meant to identify the peers of a customer using a static API key directly embedded in the customer's website. As a result, the attacker could easily retrieve the key to free-ride the PDN service, which increases the cost of the PDN customers. Although PDN providers utilize a domain whitelist to prevent the API key abuse, our experiments found that an attacker can still bypass the protection through a custom client. Even worse, many PDN customers do not even enable the whitelisting protection for the convenience of sharing API keys across multiple domains. Among the 40 API keys we extracted from the detected PDN customers, 11 disabled the domain whitelisting protection.

Another security-critical weakness we identified is video segment pollution: a malicious peer could alter any video segment it receives and forward it to other benign peers without being noticed. Although content pollution is a known threat in traditional P2P networks, previous attacks [39], [53], [54] rely on understanding of P2P protocols and access to local storage, which are not applicable to PDN scenarios since PDN services utilize customized data protocols and store the downloaded data in the cache under the protection of browsers. In our research, we proposed a novel attack wherein the attacker can pollute arbitrary video segments without knowledge of P2P protocols or access to local storage. Our evaluation results demonstrate the feasibility of the attack over all public PDN providers and a demo [1] is published online: https://sites.google.com/view/pdnsec/home/demo.

Also discovered are extensive peer IP leaks in PDN services: a PDN service automatically connects each peer to available viewers to establish P2P connections. Previous works [32], [47] have reported the exposure of real IPs to websites through WebRTC, which could be abused for web tracking. In our research, we discovered a *different yet more serious* IP leak caused by the PDN service, exposing viewers' IPs to untrusted peers. Our case studies show that two popular websites and apps, i.e., Huya TV and RT news, exposed respectively 7,055 and 685 viewers' IPs to a peer we deployed to watch a single live streaming channel 2 hours per day for just one week. Our experiments further indicate that all existing PDN services, both public and private ones, do not have sufficient protection in place to restrain viewers' IP exposure.

Furthermore, we found that PDN services consume peers' computing and bandwidth resources without consent. Among the PDN services we detected, no matter public or private, none asks for viewers' permission and viewers have no option to disable the PDN service. We further measured peers' resource consumption and found that a PDN service generally incur 15% more CPU and 10% more memory usage. Also, our experiments show that the upload traffic of peers increases significantly (up to 200% of the download traffic with 3 peers) as the number of peers grows, while their download traffic does not go up accordingly. Our further analysis on Peer5 customers reveals that 3 highly popular apps (i.e., *com.bongo.bioscope*, *com.portonics.mygp*, *com.arenacloudtv.android*) even allow the PDN service to use viewers' cellular data for both uploading and downloading, which may incur extra financial cost to viewers.

**Mitigation**. To mitigate the security risks discovered, we discuss the limitations of known defense mechanisms and present several protection suggestions along with a feasibility evaluation under a simulated environment. More specifically, for the service free riding risk, we propose an authentication mechanism that utilizes a video-binding and disposable token, which can effectively demotivate unauthorized use of the PDN service purchased by others. To address the video segment pollution threat, we propose a peer-assisted defense mechanism wherein the PDN server randomly selects a subset of peers to report and verify the integrity metadata (IM) for each video segment. This protection raises the bar for a content pollution attack, which will only succeed when all randomly selected peers are malicious. We also discuss the countermeasures for the peer

IP leak risk through limiting the candidate peers by the geolocation, ISP, etc., or deploying TURN servers to relay peer-to-peer traffic.

**Contributions**. The contributions of the paper are outlined as follows:

• *A large-scale characterization on real-world PDN participants*. Our research reveals the technical mechanisms of PDN services as well as their prevalence through an automatic detection framework, leading to the discovery of 134 websites, 38 Android apps and 9 private PDN services.

• *The first study on PDN security*. We report the first comprehensive study on PDN security based upon a novel PDN analysis framework. Our study leads to a new understanding of the security implications of today's PDN ecosystem, including its weaknesses in protecting PDN customers (*service free riding*, *video segment pollution*) and viewers (*peer IP leak*, *resource squatting*).

## 2. Background

**Peer-assisted video streaming.** Due to the significant benefit of traffic savings, peer-assisted video streaming has been adopted by numerous commercial CDNs including Xunlei Kankan [75], LiveSky [73], Spotify [45], and Akamai [77]. These P2P-CDNs generally require end users to install client-side software and design ad hoc protocols for peer-to-peer communication. It is also a key challenge to integrate P2P with existing CDN services [73]. As previous P2P-CDNs discontinued for various reasons [16], PDN services based on WebRTC emerge as the next-generation peer-assisted video streaming network, which provides a more convenient SDK and better security mechanisms. These PDN services are also embedded into web players [6] and enterprise content delivery (eCDN) [8] for wide deployment. Regarding network structures, peer-assisted networks can be classified into two types: tree-based and mesh-based [44]. Literally, tree-based networks organize peers in a structure of multiple trees, selecting some peers as root nodes and others as leaf nodes. In a mesh-based network, peers dynamically connect to a subset of random peers based on attributes such as content/network availability. In our research, PDN takes the mesh-based network.

**Video streaming protocols**. A video streaming protocol specifies how media data is delivered over the Internet. Early examples of such a protocol include Real Time Streaming Protocol (RTSP) [68] and Real Time Messaging Protocol (RTMP) [63]. However, these protocols are either proprietary or do not support video streaming through HTTP. Thus, in the past decade, a set of HTTP-based adaptive bit-rate protocols have emerged and gained popularity, particularly HTTP Live Streaming (HLS) [5] and Dynamic Adaptive Streaming over HTTP (MPEG-DASH) [70]. These protocols break a video into smaller segments that can be downloaded through HTTP. Such segments are made available at different bitrates, so as to allow the video client to adapt the video streaming to various network conditions. A manifest file is also created to trace these video segments. Among these protocols, WebRTC and Secure Reliable Transport [11] (SRT) are characterized by new features such as lower latency, security, P2P support, etc.

**WebRTC**. As mentioned earlier, PDN services today are built on top of Web Real-Time Communication, also known as *WebRTC*. WebRTC allows web browsers to communicate directly in a P2P manner. It has been supported by most modern browsers, including Mozilla Firefox 22+, Google Chrome 28+, Safari 11+, Opera 18, and Microsoft Edge 12+. It becomes a primary choice for streaming services that utilize P2P communication between their viewers to lower bandwidth consumption and reduce their CDN traffic cost. WebRTC is free and open-source and can be easily deployed using regular JavaScript APIs. It is by default built into browsers, allowing the users of the streaming service to receive its P2P support without installing any additional software. WebRTC has two main components: the *signaling component* and the *data channel component*. The former manages the communication between the application (e.g., the website) and the peer (e.g., the browser). This component acts as a medium to gather communication information between peers. The latter controls the transmission of the data between the connected peers. To establish a P2P connection between endpoints (e.g., web browsers), a specified signaling server exchanges the meta-data required for the two parties to communicate with each other. More specifically, an endpoint connects to the signaling server, typically through HTTPS, and then generates the required meta-data in two categories: 1) media preference, 2) network information. The media preferences are set using Session Description Protocol (SDP) [50], which specifies a variety of media options, such as media type. Also set is Interactive Connectivity Establishment (ICE) [52] candidates with network information, like the public IP address essential in establishing the P2P connection between two parties. In order to obtain the IP, a Network Address Translation (NAT) traversal server such as STUN or TURN is used. The SDP and ICE candidates are then relayed to the endpoint through the signaling server, typically using the WebSocket protocol [43]. Once the information is accepted, the P2P connection between the endpoints is established.

## 3. Understanding the PDN Ecosystem

In this section, we present our understandings of the PDN ecosystem. We first introduce a typical PDN scenario and its key players, i.e., PDN providers, PDN customers, and peers (§3.1). Then we describe our profiling of a set of representative PDN providers' service models and operations (§3.2). Such profiling enables us to develop techniques for detecting PDN customers (both websites and mobile apps) at a large scale (as reported in §3.3).

### 3.1. The Ecosystem

Generally, the PDN ecosystem consists of three key players, i.e., PDN providers, PDN customers, and peers.

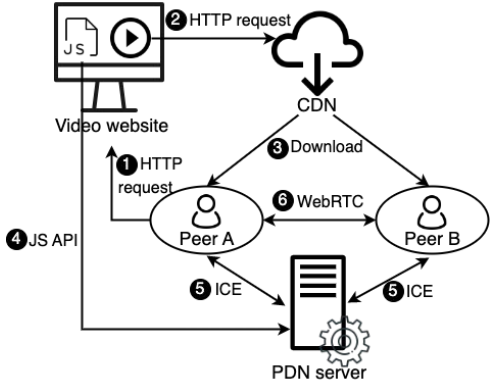

Figure 1: Traffic flows of a typical PDN scenario.

Among these players, PDN providers offer the PDN service and the integration SDK, PDN customers refer to video streaming services and their apps or websites that subscribe to the PDN service, and peers denote the video viewers of these video websites or apps. Figure 1 illustrates a typical traffic workflow in the PDN ecosystem. As claimed by PDN providers, a PDN combines the traditional CDN with WebRTC-based P2P network and offloads a large amount video traffic to P2P connections by integrating a JavaScript API into a video website. Specifically, in the traditional CDN mode, when a viewer (e.g., Peer A) opens a video website ❶, it sends an HTTP request to the CDN which stores the specified video files ❷ and downloads the video files before playing it with a video player ❸. If the PDN service is enabled, the Javascript API embedded in the video website will automatically initiate a WebRTC interface and connect to the PDN server ❹. After the process of Internet Connectivity Establishment (ICE) between peers and the PDN server ❺, Peer A shares the meta data of video files and its network information (e.g., IP and port) with other peers (e.g., Peer B). Then Peer B can request video segments from Peer A instead of the CDN network ❻.

PDN distinguishes from previous P2P-CDN services in two aspects: First, due to the wide support of WebRTC in modern browsers, PDN services do not require any dedicated client-side software and can be activated through loading a JavaScript SDK within a web browser. Moreover, PDN services are compatible with existing video streaming protocols and CDN services. However, such features make the PDN services almost unnoticeable to viewers and significantly increase the difficulties of detecting the existence of PDN services.

### 3.2. PDN Providers

Our research first identified popular PDN providers and subscribed to their services to gain an inside view of their operation and business. Then we estimated their popularity using passive DNS.

**Identifying PDN providers**. To find PDN services, we first queried Google Search with PDN-related keywords like "P2P live streaming" and "P2P CDN", and manually checked the search results. The providers identified were further verified by contacting them in the name of video streaming services to get their service details. In this way, we confirmed three popular PDN providers: Peer5 [24], Streamroot [20], and Viblast [29]. We further signed up as a customer of the verified PDN services so as to access their documentation, client-side SDKs as well as customer portals, which enabled us to gain insights into how these services work. Based on the new insights, we constructed a set of robust PDN signatures to help identify potential PDN customers, and manually verified them to understand their effectiveness (§3.3). Subscription to the services also allows us to enable the PDNs on our experimental video streaming website, through which we captured PDN traffic to study the PDN protocols and workflows (Figure 1).

**Service model and pricing policy**. All the PDN providers identified in our research claim that their services can offload at least 50% traffic from existing video streaming infrastructures to PDN peers [20], [24], [29], and will become more efficient when the traditional server infrastructures are weighed down by traffic peaks. We observed that these services have different pricing policies for their customers, based upon PDN traffic volume, concurrent video viewing hours, the number of concurrent viewers, etc. Table 1 summarize the characteristics of the PDN providers. And we can see, most providers have cross-platform SDK support covering desktop web browsers (Web), mobile devices, and over-the-top (OTT) devices. Also, customers are charged according to either the concurrent viewer hours or the traffic volume offloaded to PDN peers. Specifically, Peer5 will charge a customer $500 for offloading 50TB video traffic from the original video server. Differently, Viblast will charge customers at a rate of $0.01 for the first 10,000 concurrent viewer hours.

**Popularity estimation**. We then move to profile the popularity of these PDN services, i.e., to what extent they are adopted in real-world video streaming activities, which is non-trivial without access to PDN operations. Through the aforementioned PDN traffic analysis, we observed that each PDN provider provided its peers with a backend server uniquely identified by a domain name, e.g., *ws.peer5.com* for Peer5. And the traffic volume towards the backend server can serve as a lower-bound approximation for measuring a PDN service's popularity. We thus estimated their traffic volumes by consulting a passive DNS service (*p*DNS) (i.e., Farsight Security [21]) for their backend server. Table 2 presents the average daily estimate of the number of DNS resolutions to the three PDN providers' backend domains in the recent 4 years. From the table, we can see that Peer5 and Streamroot are the most popular PDN providers, and each receives over 20K *p*DNS requests per day. In the recent 4 years, peer traffic volume has seen a sudden decrease in 2019 and then a stable growth since then except for Peer5, possibly due to the fact that it was acquired by Microsoft in 2021 [9].

TABLE 1: Popular PDN providers

| Provider | Supported Platforms | Pricing Standard | Pricing |
|---|---|---|---|
| Peer5 | Android,iOS,Web,OTT | monthly p2p traffic | $500 for 50TB |
| Streamroot | Android,iOS,Web,OTT | monthly p2p traffic | unclear |
| Viblast | Android,iOS,Web | concurrent viewer hour (CVH) | $0.01 per CVH |

TABLE 2: pDNS resolutions of popular PDN providers

| Provider | Backend server | Average daily resolutions | | | |
|---|---|---|---|---|---|
| | | 2018 | 2019 | 2020 | 2021 |
| Peer5 | ws.peer5.com | 48,124.33 | 24,795.05 | 31,575.55 | 22,152.62 |
| Streamroot | backend.dna-delivery.com | 32,823.76 | 24,968.69 | 34,215.89 | 36,805.73 |
| Viblast | cs.viblast.com | 231.43 | 24.04 | 50.28 | 390.56 |

TABLE 3: Detected PDN customers

| PDN Provider | # Websites | # Apps | # APKs |
|---|---|---|---|
| Peer5 | 60 | 31 | 548 |
| Streamroot | 53 | 6 | 68 |
| Viblast | 21 | 1 | 11 |
| Total | 134 | 38 | 627 |

### 3.3. PDN Customers

In our research, we proposed a novel framework to detect PDN customers at a large scale, including the websites and Android apps. Specifically, we first identified websites and Android apps integrated with PDN services, which we call *PDN customers*, using a signature-based approach. Then we analyze the popularity of these *PDN customers*.

**Challenges of detecting PDN customers**. Intuitively, a straightforward method to determine whether a website or Android app enables the PDN service is to utilize the unique pattern of the peer-to-peer traffic (e.g., STUN requests, DTLS traffic between peers, etc.). However, this method turns out to be challenging and costly. As mentioned earlier, a PDN service can only be triggered when enough clients are watching the same video simultaneously. Also, due to the dynamic feature of JavaScript, PDN services will be loaded dynamically. For example, *severestudio.com* will load the PDN service only when there is an active streaming source on the page. These conditions render PDN activities less observable during dynamic analysis. Therefore, simple static signature-based detection becomes a better alternative, though this approach may still miss many cases, particularly, when the JavaScript code containing PDN signatures is heavily obfuscated or is loaded under specific conditions at runtime. So our current approach is conservative, only identifying a subset of PDN customers.

**Detecting PDN customers**. Through an analysis on PDN services (§3.2), we collected a set of robust signatures to fingerprint their customers, which were extracted from their documentation and source code (js files or mobile SDK). These signatures consist of URL patterns (e.g., *api.peer5.com/peer5.js?id=*), unique namespaces (e.g., *com.viblast.android*), and meta-data in the Android manifest file (e.g. *io.streamroot.dna.StreamrootKey*). All identified signatures are presented in Table 5 in the Appendix. Leveraging these identified signatures, we built up a scanner to crawl high-profile websites, in an attempt to find potential PDN customers.

Since PDN mainly serves video streaming platforms, we only considered the popular video-related websites and Android apps in our research. Specifically, we first queried the top 300K domains according to Tranco Top Sites Ranking [64], which turns out to be more stable and secure than other ranking lists such as Alexa. Then we collected the category information of these top domains as provided by the 5 category engines in VirusTotal, i.e., Forcepoint ThreatSeeker, Sophos, BitDefender, Comodo Valkyrie Verdict, and alphaMountain.ai. For each domain, if any of the 5 engines returns a category label containing keywords such as "tv" or "media", we consider it a video-related domain. In this way, we found 68,713 top video-related domains through this category filtering. Also, we queried source code search engines NerdyData [23] and PublicWWW [25] using PDN signatures as keywords, which reported 44 potential PDN-enabled websites. Altogether we gathered 68,757 domains for our PDN detector.

We then performed a signature-based scan between Jan 2022 and Feb 2022 using Selenium [12], a framework for automatic web application testing. Our scanner dynamically crawls the website of a given domain by downloading its HTML files and all JavaScript files if the site contains a "video" tag on its web page, and then traverse all the subpages under the same domain until a PDN signature is found. To limit the depth of searching, our scanner only examines the subpages within a depth of 3. To avoid non-negligible overhead to a website, we limit the crawl rate to 1 webpage per 3 seconds with a timeout of 10 minutes for a given domain. If any PDN signature is found in these subpages, the scanner considers the domain as a *PDN customer*.

We also collected popular apps and their APKs from Androzoo [33], a large repository of Android APKs from multiple app stores, including Google Play, Anzhi, and AppChina. By June 2022, it contains 19,661,675 different Android APKs from 7,954,395 apps. Since Androzoo does not provide information about the app category or downloads, we randomly sampled 1.5M apps among the 8M apps. Our scanner automatically downloaded the latest APK version of sampled apps and then unpacked it to search PDN signatures on Android, as illustrated in Table 5 in the Appendix. An APK is considered as a *PDN customer* if it contains at least one PDN signature. We further checked all historical APK versions of detected apps to estimate the scale of different APK versions.

**Popularity of PDN customers.** Our study has led to the discovery of 134 websites and 38 apps (with 627 different APK versions) as PDN customers. As shown in Table 3, among the services behind these customers, Peer5 is the most popular one, with 60 websites and 31 Android apps, followed by Streamroot with 53 websites and 6 Android apps. The other 21 websites and 1 Andorid app were detected to be integrated with the Viblast PDN SDK. Furthermore, most PDN customers are found to have a large number of viewers. To measure the popularity of PDN customers, we further queried the monthly visits of PDN websites from SimilarWeb [28] and the installs of PDN apps from Google Play. Among the 134 detected websites and 38 apps, we successfully obtained the data of 105 websites and 35 apps, and the others are not found in SimilarWeb or Google Play.

From our results, 92 (69%) PDN websites are visited over 1 million times per month, and 19 of them have over 10 million monthly visits, including popular websites *www.rt.com*, *www.clarin.com* and *www.rtve.es*. Among the 38 Android apps, 25 (66%) of them have been downloaded by over 1 million times and 9 have more than 10 million downloads, including 1 app (*com.graymatrix.did*) with over 100 million downloads, and 1 (*iflix.play*) with over 50 million downloads. We list the top detected PDN websites and apps in Table 6 and Table 7 in the Appendix. Our measurement indicates that PDN services have been adopted by many popular customers and have reached millions of video viewers.

**Private PDN services.** We also observed that some websites are embedded with HTML or JavaScript code that shares similar patterns with ones from PDN services, but they are not customers of either of the three PDN providers nonetheless. Unlike known PDN providers, these cases do not involve third-party API keys or external Javascript APIs, but associate their own domains (usually their subdomains or relevant domains) with the involved PDN servers. Thus we consider them as *private PDN services* since they are ad hoc services with each dedicated to a specific video/live streaming platform.

Specifically, our detector identified 385 websites matching the general WebRTC-related signatures (see Table 5 in the Appendix). We manually studied the traffic of the top 57 websites that rank in top 10K websites, and confirmed that 9 has integrated proprietary PDN functionalities (i.e., private PDN services), including 5 popular video streaming platforms, e.g., Youku (*youku.com*), Tencent Video (*v.qq.com*), and 4 live streaming platforms, e.g., OK Social Network (*ok.ru*), Huya (*huya.com*). We list these popular private PDN services in Table 8 in the Appendix. It is interesting that such private PDN services are extremely popular in China, covering most top video hosting and live streaming platforms. We also noticed that 2 adult video platforms, i.e. *xhamsterlive.com* and *stripchat.com*, utilize WebRTC protocols to relay traffic. For the other 46 cases, we confirmed 3 cases invoking WebRTC APIs for web tracking. Yet for the most cases, we were unable to determine why they contain such signatures, since we failed to trigger any WebRTC-related traffic. Even for those confirmed PDN customers, it is really hard to trigger the peer-to-peer traffic. Some websites (e.g., *youku.com*, *tudou.com*) have restrictions on the viewers' locations due to copyright, and some websites (e.g., *younow.com*) require us to register an account to view the content. All these conditions limit the scale of our analysis on these private PDN customers. Also, these private PDN services mainly serve the viewers of their domains only and are deeply embedded in their HTML code, making it impossible to integrate these services on our own test website. Due to these factors and ethical concerns, we do not evaluate them for the service free riding risk (§4.2) and the video segment pollution attack (§4.3).

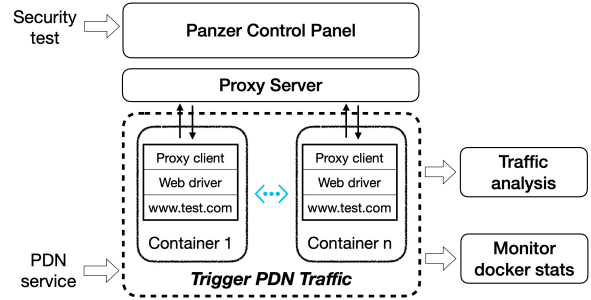

Figure 2: The architecture of the PDN Analyzer

## 4. Security Risks in PDN

With the in-depth understanding of the PDN ecosystem (§3), we move forward to reason about and analyze potential security risks introduced by PDN services to the parties involved, especially the video viewers and PDN customers. Our exploration is based upon the observation that PDNs have a significant impact on the original threat model for video streaming where video content is directly offered by trusted parties (e.g., CDN) and distributed through secure channels (e.g., HTTPS). Following, we first present a PDN analysis framework and then elaborate on four security risks discovered in our research and their impacts.

### 4.1. PDN Analyzer

In order to reliably and effectively test the potential risks imposed by PDNs, we developed an automatic PDN analysis framework, as illustrated in Figure 2. At a high level, our PDN analyzer accepts a PDN service and an execution plan as the input, which specifies what peers do and how to do. Then, it runs each PDN peer as a separate Docker container equipped with a web driver and a proxy client, which communicates with a configured proxy server controlled by our PDN analyzer. Once the execution finishes, our PDN analyzer returns a dumped network traffic report as well as all execution logs, which can be analyzed to decide whether the risk under evaluation is triggered. Note that all the security tests and log analysis are automatically performed by our PDN analyzer. To simulate the PDN service in the real world, we integrate PDN services on our own website (www.test.com) and a customized stream server connected to a CDN service. Specifically, we rent an AWS EC2 instance with Wowza Streaming Engine deployed and set up our own video streaming source. And we utilize Amazon CloudFront as our CDN service to distribute our video content.

**Threat model**. We assume the attacker is able to participate in the PDN system as a PDN peer, as well as intercept the TLS traffic from/to the peer under its direct control. To transparently intercept the TLS traffic of the peer under its control, the attacker can configure the peer with a self-signed root certificate. Note that we do not require the attacker has any knowledge of PDN implementations or protocols, nor does the attacker need to access the local storage, i.e., the cache of browsers.

**Triggering PDN traffic**. In the PDN analysis framework, PDN peers run as separate containers to perform automatic webpage rendering and video playing. To this end, we use Selenium WebDriver [12], a standard browser automation framework, to control a browser (Chrome by default) through its driver (e.g, ChromeDriver [2] for Chrome, and geckodriver [4] for Firefox). Specifically, our PDN analyzer first creates a container with parameters specified in its execution plan. Inside the container, a browser is then instructed, through a Selenium script, to automatically render a given video streaming URL sourced from the execution plan. Once the webpage is rendered, the next step is to play the video embedded in the webpage so PDN traffic can be triggered. Here a few tricks are deployed to make the video playing automatic. Specifically, a command-line parameter of *"–autoplay-policy=no-user-gesture-required"* is specified when starting the Chrome browser, which enables video autoplay without user interactions. Further, *"autoplay=true"* is appended to the target webpage URL as a parameter, which is a common practice for instructing a video player to automatically play a given video [31]. Besides, network traffic of different containers is mapped to different ports to avoid interference. Our experiments show that PDN traffic can be reliably triggered when running multiple PDN peers through these automatic steps.

**Monitoring PDN activities**. When running PDN peers as separate containers, we want to monitor PDN activities in terms of network traffic and resource consumption (e.g., CPU and memory). To monitor the network traffic, tcpdump [13] is started to dump incoming and outgoing network traffic on the default virtual network interface *docker0*, following the creation of the parent container. Furthermore, the PDN analyzer utilizes Docker Engine APIs [3] to monitor in real-time the container stats especially the resource consumption stats such as CPU usage, memory usage, and network I/O.

**Security risks in PDN.** We attempted to uncover security risks in PDN through PDN documentation reading, taking lessons from previous studies on P2P and peer-assisted video streaming [34], [44], [77], and preliminary experiments. As a result, multiple concerning security risks are identified, i.e., *service free riding*, *video segment pollution*, *peer IP leak* and *resource squatting*. More specifically, we found that a malicious video website or app can free-ride legitimate PDN services (*service free riding*); a malicious peer is able to pollute video segments of benign peers (*video segment pollution*). For the risk of *peer IP leak*, we identified extensive leaks of viewers' IP addresses. Also, all of the identified PDN services consume extra viewer resources (CPU, memory, bandwidth etc.) without consent (*resource squatting*). We even found 3 popular Android apps which were configured to consume cellular data for both uploading and downloading traffic in PDN. Following, we detail these security risks and their security impacts.

**Ethical considerations**. We carefully designed our methodology to minimize any real-world ethical impact. Specifically, we applied for IRB approval from our institution and performed all our experiments under the received guidelines. For experiments requiring PDN access, we gained permissions from PDN providers. Also, these experiments were run on our own test website integrated with free trials of PDN services, which would not affect any real-world viewers or PDN customers. In the peer IP leak test, we only collected IPs of viewers connecting with our controlled peer. Also, we focus on measuring the coarse-grained geographical distribution of PDN peers, and have deleted the raw IP addresses given the statistical results are extracted.

## 4.2. Service Free Riding

As discussed in §3.2, public PDN services operate in a pay-as-you-go model and a PDN customer is charged for every use under its name. Leveraging our PDN analyzer and subscriptions to PDN services, we explored whether the use of a PDN service is well authenticated and whether a PDN customer can be overcharged for the use incurred by other parties. It turns out that a persistent access token (API key) issued by the PDN provider is used to authenticate PDN customers and PDN peers. Such an access token was found to be publicly visible to attackers since it is statically embedded by the PDN customers in either the PDN mobile app or the video webpages. This allows an attacker (e.g., a misbehaving video streaming site) to easily steal a legitimate PDN customer token through either a colluding peer or static analysis of the respective mobile app if available. It can later utilize the token to *free-ride* the PDN service at the cost of the legitimate customer, or even maliciously consume P2P traffic between controlled peers to incur extra cost to the targeted PDN customers. Our evaluation of real-world customers further confirms the pervasiveness of the service free riding risk. Following we elaborate on the service free riding risk and our findings, under the threat model where an attacker is capable of retrieving the PDN access token and the domains (origins) from a legitimate PDN customer.

**Service free riding**. We found that the service free riding risk is *inherent* to today's public PDN services due to the way they operate. As elaborated in §3.2, a PDN service is meant to manage the interactions among the viewers of its customer. Since these viewers are not known to the service in advance, their association with the customer needs to be proven to the PDN service. Serving this purpose is the customer's access token, which many services use for authenticating viewers of their customers to the service through the JavaScript code dispatched to the viewers' browsers. In the absence of protection, however, such tokens can be easily retrieved from a PDN customer's website or mobile apps for impersonating legitimate customers and their viewers.

We then performed an attack feasibility test within our PDN analyzer framework. Specifically, we first integrate a PDN SDK into our own test website (www.test.com) with the API key extracted from another PDN customer. Then we run two peer containers and configure the web driver to watch the same video stream with each peer. We further analyze the traffic between these containers (peers) and the PDN server: an initialization request (either accepted or rejected) from a peer to the PDN server indicates that the API key is valid, and a successful binding of the peers provides evidence that the API key can be abused and thus the service free riding risk exists. During the experiment, we disabled the auto-play function and ensured that no data was actually transferred between peers, thus no cost was generated for customers owning these tokens.

**Evaluation of service free riding**. In our study, we evaluated how real-world PDN customers protect their PDN subscriptions against the service free riding attack. Among the PDN customers of 134 websites and 38 apps, we successfully extracted out 44 API keys, despite heavy obfuscation and dynamic loading as enforced by some PDN customers. Among these API keys, 36 are from Peer5, along with 7 from Viblast and 1 from Streamroot. We tested all of them and found 40 were valid during our test. Among the 40 valid API keys, 11 were confirmed to be vulnerable to the service free riding attack and all of them belong to Peer5. The 11 API keys have been integrated by 18 PDN customers and 4 of them are shared by multiple domains or apps. We also inspected the default settings of all the three PDN providers. It turns out that domain whitelisting is disabled by default for Peer5 and Streamroot, while Viblast requires setting up the whitelist when enabling the service.

To verify the effectiveness of domain whitelisting protection, we further evaluated the service free riding attack with domain whitelisting enabled. We applied for a free trial from all the 3 PDN providers and succeeded for Peer5 and Viblast, while Streamroot declined our request. In our settings, we integrate the PDN service into our test website (www.test.com) with the trial API key. And we set a domain whitelist to include a specified domain (www.example.com).When a viewer visits the victim website (www.example.com), the proxy server redirects the traffic to the target website (www.test.com), which decieves the viewer client to initiate a legitimate request originating from the domain (www.example.com) and send it to the PDN server. We then decide whether peers of our test website can circumvent the domain whitelisting protection by observing the response from the PDN server. As a result, both Peer5 and Viblast were found to be vulnerable to such domain spoofing attack. Although we were unable to test Streamroot service, it is susceptible to our attack as it also relies on a static API key and domain whitelisting to authenticate its customers [26].

### 4.3. Video Segment Pollution

Different from existing P2P networks (e.g., BitTorrent), PDN enforces protection mechanisms over both the commu-nication channels and the storage. First, PDN utilizes TLS encryption to protect ICE communication between peers and the PDN server (❺ in Figure 1). Also, peers in PDN are connected via WebRTC, which supports video streaming protocols over DTLS encryption (❻ in Figure 1). Second, PDN caches the downloaded content in the memory of browsers, which is protected by the same-origin policy and purged after a short time. Such protections render previous content pollution attacks ineffective. Thus PDN providers claim the PDN service is as secure as traditional CDN services [27]. In our research, however, we proposed a novel attack to compromise the content integrity in PDN. Our attack is based on the observation that the PDN server is unable to verify whether a video file is downloaded from the original CDN. Although the other channels are well protected under the assumptions, the attacker can still download "fake" video segment files and then spread them to other peers with the help of a malicious peer. Note we assume that the attacker has access to the original video files and the corresponding meta files. This is practical with the help of existing browser plugins such as *Live Stream Downloader* [7].

TABLE 4: Security risks of PDN services

| Security Risks | Peer5 | Streamroot | Viblast | Private |
|---|---|---|---|---|
| Service free riding | ✓ | – | ✓ | – |
| Video segment pollution | ✓ | – | ✓ | – |
| Peer IP leak | ✓ | ✓ | ✓ | ✓ |
| Resource squatting | ✓ | ✓ | ✓ | ✓ |

**Our attack**. Our idea is to run the proxy server in the middle of a controlled peer and the real CDN. The proxy acts as a fake CDN to download video files from the real CDN, automatically replaces the video files before forwarding to the controlled peer. As illustrated in Figure 3, the proxy client of the attack peer is configured to redirect the CDN URL to the proxy server ❶ when the attack peer visits a video website, the video source URL (pointing to the CDN that stores the target video) is utilized by the proxy ❷ to download the original video files for alteration ❸, and then store the altered video files to an attached fake CDN ❹. Then the HTTPS proxy redirects the source URL to the fake CDN for downloading the altered video ❺. When the attack peer plays the altered video, it deceives the PDN server and other peers that the attack peer is watching the original video content, allowing the attacker to propagate the video to other victim peers ❻. In our test, if we observe consistent data communication sending from the attack peer to other victim peers over the DTLS channel, it means our attack succeeds and the content integrity is compromised.

As we discussed in §2, video streaming protocols usually split a large video file into small segments and a manifest file (e.g., an M3U playlist) is utilized to track these segments. In the PDN analysis framework, we first run the *naive content pollution test*, in which the attacker directly replaces video segments and the corresponding manifest file. However, we found PDN services would detect such simple pollution attack and prevent the malicious peer to connect with other peers. This is possibly because PDN servers utilize the manifest file to distinguish video sources and decide which viewers can serve as peers for a given video. Therefore, changes in the manifest file will make a video viewer excluded from peer-to-peer video segment transmission. To bypass this constraint, we designed the *video segment pollution test*, in which the attacker replaces one or more video segments while keeping the original manifest file unchanged. We detailed our method of generating polluted video segments in the Appendix 9.2.

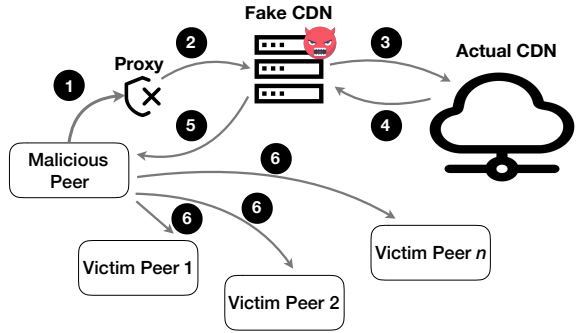

Figure 3: Illustration of our attack

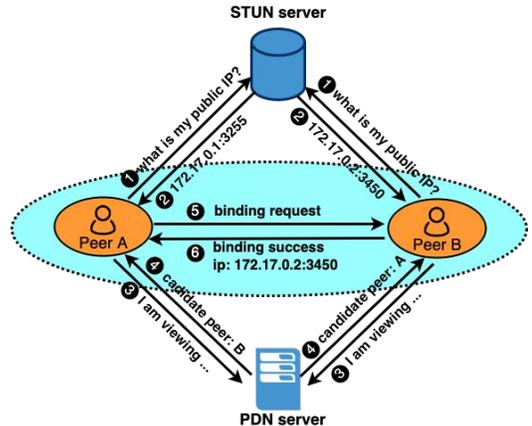

Figure 4: Illustration of IP exchange in PDN

**Test results**. As mentioned in §4.2, we leveraged free-trial access to Peer5 and Viblast and integrated their PDN SDKs into our test website. In our settings, we configure the web driver to open our test website on two different peer containers and watch the same video content simultaneously to make sure these two viewers establish a peer-to-peer connection. As a result, both Peer5 and Viblast failed to pass our *video segment pollution* test, which demonstrates that the attacker could utilize our method to pollute video segments of a victim PDN customer. We also published a video demo ( https://sites.google.com/view/pdnsec/home/demo) to illustrate the video segment pollution attack. From our test results, we inferred that PDN services probably utilize the manifest file to verify the integrity of a video stream. However, it turns out to be vulnerable and can be easily bypassed through our method. As revealed in a recent study [71], a content pollution attack in a P2P live streaming system will quickly propagate to 47% of viewers in the initial stage even when the initial number of polluters is small. Considering the popularity of PDN customers, the video segment pollution attack in PDN can be very destructive and may impact millions of viewers.

### 4.4. Peer IP Leak

It has been reported that WebRTC can be abused to cause unauthorized IP and port leaks [32], [40], [47]. A website or app can easily exploit the WebRTC API to access viewers' real public IP addresses and ports through a STUN server, even if the viewers are hidden behind a VPN. It can be used by websites or apps for tracking users' activities [40]. On the PDN service, however, we have observed a *different yet even more serious* IP leak risk, which exposes viewers' real IPs to other peers and thus made it possible for the attacker to harvest these IPs as a peer. To seek the video segment from other viewers, a peer automatically requests from the PDN server a list of candidate IP addresses of those watching the same video. Then the peer tries to exchange IP information with all candidate peers to test the connections and find out the most cost-effective source for downloading the video. During this process, inevitably one's IP and port are disclosed to other video viewers, even though they do not know each other and would not trust each other.

Figure 4 illustrates the detailed IP exchange process in the PDN. During the process, a video viewer first sends a STUN request to a public STUN server ❶ which responds with the viewer's real public IPs and ports ❷. Following, it tells the PDN server what video stream it requests ❸, and as a response, the PDN server replies with a list of available candidate peers selected upon the content they are watching ❹. Then the viewer can send a binding request to each of the candidate peers ❺ which in turn responds with its IP if the binding request is accepted ❻. The whole process will be performed every few seconds to check the liveness of peers. Moreover, we find the binding success response ❻ containing the real public IP address and port is transmitted in plain text, which further exaggerates the risk of IP exposure. In the absence of proper protection, this exposure can be very extensive, not limited to those geographically close, as demonstrated below.

**Peer IP leak in the wild**. To measure the extent of peer IP leaks in PDN, we utilized our PDN analyzer to collect and analyze peer IPs from popular PDN customers. In our test, we chose two popular PDN customers: RT News (*com.rt.mobile.english*), a mobile app integrated with Streamroot SDK, and Huya TV (*huya.com*), a website with a private PDN service. To minimize ethical risks, our tests collected only the IP addresses of PDN peers communicating with our controlled peer, which were removed immediately after generating the aggregated statistics. Specifically, we collected two-hour traffic from a controlled peer in a live channel for each of the two customers lasting for 7 days. From the traffic, our PDN analyzer extracted the STUN packets for exchanging IP addresses with our peer (i.e., ❺ and ❻ in Figure 4) and recovered the IP addresses of the candidate viewers trying to establish a peer-to-peer connection. Altogether, our PDN analyzer gathered 7,740 unique peer IP addresses, including 7,055 from Huya TV and 685 from RT News. We then further queried IPInfo [22] for these addresses' geolocations and other information and found that 7,159 of these IPs are public IPs, along with 581 as bogons [19]. Among these bogon IPs, 543 are in private networks, 33 are for NAT [72]), and the other 5 are reserved IPs. These IPs (private, NAT, reserved) were returned probably due to the errors in the NAT traversal process, which replied with unreachable IPs to our controlled peer. Also, among the public IPs, 98% are from Huya TV in China, while IPs from RT News distribute across 259 cities in 56 countries, with United States (35%), Britain (17%), and Canada (13%) being the top 3 countries. The results are consistent with the distribution of viewers from these two PDN customers. For Peer5 and Viblast, we performed similar experiments with two controlled viewers (one located in the US and the other in China) on our test website and confirmed that two peers were able to collect the other's IP even without playing the video. Our experiments demonstrate that all PDN services expose viewers' real IPs to globally distributed audience with few protections, which suggests a reliable channel for the attacker to harvest active viewers' IPs. This potentially enables an attacker to extensively gather peers' IPs and link them to the content of the videos being watched. Further, active fingerprinting can be run on these IPs to identify and target known vulnerabilities.

### 4.5. Resource Squatting

PDN services utilize video-viewing peers for video segments encryption/decryption and transmission, which inevitably incurs computation and communication burden to peers. Traditional P2P network usually allows users to limit the upload speed and provides incentives by awarding cooperative peers. In our research, we study how significant the burden to the peers, as incurred by PDN services, could be, and whether viewers are well aware of and able to control such burden.

**User consent**. A previous study involving large-scale users [77] reveals that only around 30% of all video viewers opt-in to participate in P2P video streaming networks. Thus it is significant to ask for consents when recruiting a video viewer into PDN, otherwise, it is a compromise of privacy. To find out whether PDN customers have informed viewers of their participation in content delivery, we manually checked all the PDN customers (including the 134 websites, 38 Android apps, and 9 private cases) detected in our study and manually inspected their services and public documentations. The results show that none of them provide any pop-up windows to ask for viewers' consent or communicate with their viewers the P2P network they are about to join through "Terms of Use" or other web content. Therefore, we believe that their viewers are completely left in the dark about the price, both in terms of security risks and extra resource consumption, they are about to pay for the visit to the streaming sites. Also, none of the PDN providers we studied allow viewers to turn off the PDN function, and instead, these PDN services are active as long as viewers are watching the video content.

**Resource squatting.** We further estimate the resource consumed for supporting PDN services. In our PDN analyzer, we run a set of peer containers and configure their web drivers to open our test website simultaneously. On top of these containers, the monitor records through Docker Engine APIs the status of each container per second, including the CPU usage, memory statics and network I/O. As a comparison test, we also monitor the overheads when peers open a website without PDN services.

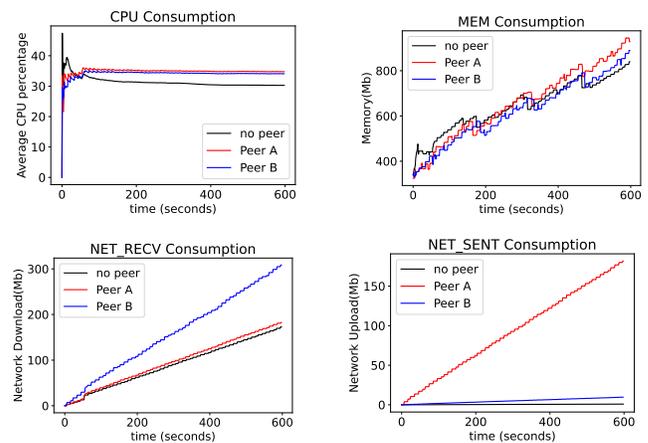

Figure 5: Resource consumption of PDN service

Figure 5 shows our test results on the Peer5 PDN service, including the CPU and memory usage and download/upload bytes measured under two peers, Peer A and Peer B, together with *no peer*, which means viewers directly requested the video from CDN. As we can see, the utilization of the Peer5 PDN service incurs non-negligible overhead for both peers, at a cost of an additional 15% CPU and 10% memory. This is mainly caused during the process of data encryption and decryption to transmit the video segments. Also, Peer A uploads much more data to Peer B than it downloads from it, indicating Peer A as a *seeder*.

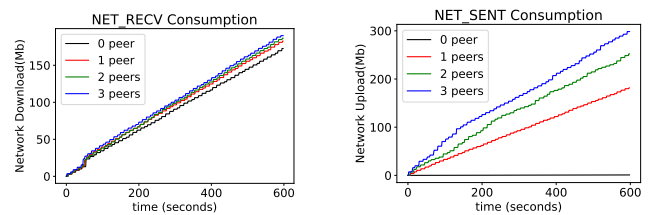

Figure 6: Network consumption of Peer A with multiple peers

We also measured the resource consumption of Peer A with the existence of multiple peers. When adding more peers up to 3 peers (other than Peer A), we found that the CPU, memory, and download traffic do not have significant differences, mainly due to the scalability of WebRTC protocols. However, the upload traffic increases significantly (up to 200% of the download traffic with 3 peers) as the number of peers grows, as shown in Figure 6. Our tests on the Viblast PDN service also present similar results. Such results indicate PDN services consume a large amount of uploading bandwidth as the number of peers grows, which is consistent with previous works [35].

**Settings of Peer5 customers** We further analyzed how real-world streaming websites configure the PDN service running in their viewers' browsers. This has been made possible by our discovery that Peer5 includes in its JavaScript code an unprotected variable specifying the configurations set by its customers (video streaming websites). Listing 1 is an example of the default settings of Peer5. Here the *deployment* parameter means the percentage of viewers to enable PDN, *mobile* means whether to enable mobile traffic for PDN, and *cellular* represents the mode to use cellular data. For the *cellular* mode, *leech* mode means cellular data is only used for downloading, *full* mode allows for both uploading and downloading, while *disable* mode does not consume any cellular data.

Listing 1: Default configuration of Peer5

```
var p5_flags = {
    "deployment":100,"page":"",
    "mobile":"enable","cellular":"leech"
};
```

We analyzed the configurations of 36 Peer5 API keys extracted from 47 PDN customers detected by our scanner (§3.3). Our study found that among the 47 Peer5 customers, 33 customers enable PDN for all the viewers, and 14 customers disable PDN by setting the deployment parameter to '0'. Among the 33 customers with PDN enabled, we identified 3 Android apps (i.e., *com.bingo.bioscope*, *com.portonics.mygp*, *com.arenacloudtv.android*) allow the use of cellular data for both uploading and downloading, and the other 30 are in *leech* mode, i.e., consuming cellular data for downloading only. As shown from our resource consumption tests, such settings may increase traffic consumption for their viewers and generate extra costs. Our measurement shows that some PDN customers consume peers' cellular data for uploading data, and 3 popular Android apps (with over 15 million Google Play downloads in total) are under such configuration, which may generate extra cellular data cost to their viewers.

## 5. Risk Mitigation Suggestions

In this section, we provide several suggestions for mitigating the risks of service free riding (§5.1), video segment pollution (§5.2), and peer IP leak (§5.3).

### 5.1. Mitigating Service Free Riding Risk

To mitigate the free riding risk, we first discuss the limitations of the general authentication mechanism in the PDN scenario, and suggest a solution utilizing a disposable and peer-binding authentication token.

**Authentication mechanism**. OAuth [46] is a widely used authentication framework proposed to authorize third-party access without providing credentials. Compared with a persistent API key, OAuth can help reduce the risk of exposing the credentials (API keys). Applying OAuth to the PDN scenario, the viewer client first requests the video website server to obtain a valid temporary token and then accesses the PDN server with the token. When a viewer sends a request to the PDN server, the PDN server will query the video website server to verify the validness of the token, which causes extra overhead to the video website server. To reduce the communications between the PDN server and the video website server, we suggest they perform a key exchange in advance. The video website server will sign the token delivered to valid viewers, allowing the PDN server to verify the integrity of the token.

However, for the service free riding risk, a malicious app can perform a man-in-the-middle (MITM) attack to redirect viewers' requests to a legitimate PDN customer and get valid tokens to access the PDN service. To defend against the MITM attack, existing mechanisms (e.g. Token binding [49]) are based on the trust of clients, which is unfortunately not the case for the service free riding attack. To address this challenge, we suggest a solution by binding tokens to valid video streams on legitimate PDN customers. In this case, the attacker cannot utilize these tokens to offload the traffic of its own video streaming, which significantly reduces the economic motivations for a service free riding attack. Also, we suggest to add usage time limits and TTL in a token to prevent a replay attack.

Listing 2: Token structure

```
"customer_id": "xx.yy",
"pdn_peer_id": "1",
"video_ids": [
    "https://xx.yy/zz.m3u8",
    "https://xx.yy/hh.m3u8"
],
"timestamp": 1619814238,
"ttl": 60,
"usage_limit": 1
```

**Disposable and video-binding token**. Listing 2 illustrates an example of the token with multiple fields, implementing the suggestions above. Firstly, *customer_id* is a string designed for uniquely identifying each PDN customer, which should be assigned from the PDN provider. Then, the PDN customer server can assign each PDN peer a unique identifier, namely, *pdn_peer_id*. Also included is a *video_ids* field to identify the set of videos to be streamed in the current page. There are multiple options to compose a video identifier, e.g., utilizing the full-qualified video URL.

Following are the token issuance time (a Unix timestamp) and a *ttl* field to denote the time to live in seconds since issuance. The *timestamp* and *ttl* jointly decide whether the token is expired or not. Another field *usage_limit* is defined to constrain the number of usage limit for this token.

To implement this token, we use JSON Web Token (JWT) [51], an open industry standard for authentication, to transmit the token and its digital signature. In our experiment, the example token in Listing 2 along with its HMAC-SHA256 signature will result in a encoded JWT of 283 bytes. Our evaluation shows that this defense incurs acceptable overhead during generation, transmission and verification, which is aligned with previous works on JWT applications [65], [69].

## 5.2. Mitigating Video Segment Pollution Risk

Previous studies [39], [55], [61], [71] have evaluated various defense mechanisms against the content pollution attack and hash-based chunk signature is considered the most promising solution. However, such protection requires the video source to sign the integrity metadata and to distribute it through the network, which is not applicable for PDN. As a plugin for the existing CDN service, the PDN server has no control of the video source or the CDN infrastructures. Although it is possible for the PDN server to download content from CDN, such behavior will incur high overhead to both the CDN and the PDN server. Inspired by previous works on comparison-based diagnosis [78], we suggest a solution of peer-assisted content integrity checking.

**Peer-assisted integrity checking.** Our peer-assisted integrity checking mechanism utilizes a similar strategy to [78] by randomly selecting PDN peers to calculate the integrity metadata (IM, e.g., cryptographic hash) for a video segment. However, our mechanism differs by utilizing a trusted PDN server, which are used to resolve IM conflicts and blacklist malicious peers. In our mechanisms, these calculated IMs are reported to the PDN server, and the PDN server consider an IM is *authentic* if all selected peers report the same IM. Since malicious peers can report fake IMs, the PDN server downloads the specified video segment from CDN and calculate the authentic IM if an IM conflict was detected. Peers reporting falsified IMs will be blacklisted. *As long as there are benign peers reporting IMs, the peer-assisted integrity checking can help identify the authentic one.* The authentic IM will be further signed by the PDN server, resulting in signed integrity metadata (SIM). The SIM for each video segment will be distributed to peers for integrity checking. Note a peer will report the IM only when the respective video segment is downloaded directly from the CDN, and a video segment downloaded from other peers must be verified using its SIM.

To ensure the integrity of video content, the integrity metadata should be robust to the replay attack. Specifically, an attacker may record a legitimate video segment along with its SIM, masquerade it as another video segment, send it to the victim peers, and thus disrupt the video delivery. Also, such a replay attack can occur not only for the same video but also across videos. Therefore, the IM should also be able to verify which video a segment belongs to as well as its position in the manifest file. In our design, the IM is calculated as the cryptographic hash of the tuple of video segment content, the video identifier, and the position of the video segment in the manifest file.

**The peer blacklist.** Since the PDN server will download a video segment from CDN when a conflict occurs, the attacker may keep sending fake IMs to increase the server overhead and traffic cost for the CDN. It is necessary for the PDN server to track peers and maintain a peer blacklist. Specifically, the PDN server assigns a unique ID to each peer at the start of the session. The ID should bind to the peer's public IP address and port or other information for tracking. If a peer is detected to have involved in suspicious behaviors (e.g., sending a fake IM), it will be blacklisted and removed from the peer candidates. Note this ID should only be visible to the PDN server in case other peers may abuse the ID for tracking.

Our evaluation in simulated environment shows that our peer-assisted integrity checking incurs negligible CPU and memory consumption and an extra latency of less than 80 ms, as detailed in Appendix 9.3.

## 5.3. Mitigating Peer IP Leak Risk

Here we also discuss potential countermeasures for PDN providers to restrain the peer IP leaks in PDN. As discussed in §4.4, a viewer's IP is exposed to all its candidate peers. A straightforward solution is to limit the number of candidate peers. Specifically, the PDN server can retrieve the real IP of peers and query the information of these IPs such as geolocation and ISP. Based on the information, the PDN server can configure candidate peers to those sharing the same country or ISP. Such countermeasure can prevent unnecessary IP exposure effectively. From the test results in §4.4, the number of leaked peer IPs will decrease significantly, i.e., only 35% leaked IPs from RT News are in the same country as our controlled peer, and none of the leaked IPs from Huya TV will be visible to our controlled peer.

Although the heuristic method above mitigates the IP exposure to some extent, an attacker can still bypass this defense through a proxy peer. Also, constraining the number of candidate peers may affect the QoS of PDN services. A fundamental solution provided by WebRTC is to relay traffic between peers through TURN servers [15]. TURN servers act as proxy servers between peers and can be utilized to circumvent network censorship [36]. With the existence of TURN servers, peers do not communicate directly and thus prevent the peer IP leak risk. As mentioned in §3.3, we observed two adult video platforms (*xhamsterlive.com* and *stripchat.com*) utilized TURN servers to relay traffic. This is probably designed to protect the viewers' privacy since watching adult videos is privacy-sensitive. However, peer communications in PDN can incur a large volume of network traffic and thus cause huge overhead to TURN servers, which is not feasible in a large-scale PDN system.

# 6. Discussion

**Limitations**. Our detection for PDN providers (§3.2) may miss ones that are either proprietary or of low public visibility, which also applies to the detection for PDN customers (§3.3). Meanwhile, due to challenges in triggering PDN traffic, we take a signature-based PDN customer detector, which is not robust to code obfuscation. It is also possible that some PDN customers that match our signatures may not enable the PDN service. Also, while the PDN ecosystem is found to be vulnerable to serious security risks, we failed to evaluate some of our security tests on the Streamroot and private PDN services. It is still not fully understood whether the free riding risk and video segment pollution attack can be exploited in these PDN services. Without access to PDN servers, it is impossible to detect whether these risks have been exploited by attackers in real-world PDN activities.

**Future works**. We plan to seek collaboration with PDN providers and explore how to detect real-world attacks targeting the PDN ecosystem as well as evaluating the proposed defense solutions (as detailed in §5). Furthermore, considering the wide adoption of private PDN infrastructure by popular video streaming services (e.g., Youku and YouNow), another future direction is to cooperate with these private PDN customers and fully test the potential risks of their service with our PDN analysis framework.

**Responsible Disclosure** We have responsibly reported the aforementioned security risks to relevant PDN providers, including Peer5, Viblast, and Streamroot. Except for Streamroot, both Peer5 and Viblast have responded to our disclosure and acknowledged the disclosed risks. Specifically, for the service free riding risk, Peer5 acknowledged that non-browser clients could spoof the origin and incur extra costs to the customers. And regarding the video segment pollution attack, both Peer5 and Viblast acknowledged the security vulnerability. Peer5 also claimed that they provide premium features to check the integrity of video segments, which need to be integrated into the customer's HTTP delivery. While Viblast mentioned that they provide a player plugin to implement an MD5 segment hash provider, which downloads the video segment from the streaming server and computes the MD5 value of each video segment for peers to compare with. In terms of user consent, both Peer5 and Viblast argue that they suggest their customers inform users of the potential resource consumption and not to use cellular traffic for uploading.

**Data and code release**. Relevant datasets and source code will be released online [1]. We will open-source most of our study infrastructure, including our PDN customer detector and PDN analysis framework. Also, the datasets produced by our study will be released, including the full list of identified PDN customers.

# 7. Related Works

**P2P video streaming network**. Multiple large-scale measurements have been explored to leverage residential peers for video streaming services, called P2P-CDNs, including Xunlei Kankan [75], LiveSky [73], Akamai [77] and Spotify [45]. All these P2P-CDNs require users to install client-side software and user consent to enable P2P services. Another set of work explored utilizing residential gateway devices such as Wi-Fi hotspots and cellular base stations [56], [57]. More recent research aims to get rid of client-side software or devices through WebRTC. Typical examples include Hive.js [67] and Maygh [76], which are similar to the paradigm under our study but not compatible with existing CDN infrastructures. Our research focuses on the emerging PDN ecosystem, which differs from P2P file-sharing networks and P2P-CDNs as it is integrated as a convenient JavaScript SDK to video websites/apps and combines the existing CDN service with WebRTC channels.

**Content pollution in P2P networks.** Content pollution attack [39], [53], [54] has been proposed in P2P live streaming networks and other P2P file-sharing systems. Prithula Dhungel etc. [39] performed the first content pollution attack in a commercial P2P live stream by mixing bogus chunks to degrade the quality of a video stream. To address the content pollution risk, a lot of works [55], [61], [71], [78] model the impact of the content pollution attack and propose defense mechanisms to mitigate the risk. Roverli P. Ziwich [78] proposed a distributed diagnosis of content pollution in P2P live streaming networks based on a comparison among all neighboring peers. Haizhou Wang etc. [71] further investigated the propagation of a content pollution attack. Our work propose a novel attack of video segment pollution in PDN, which has never been investigated before.

**WebRTC security**. WebRTC has been studied from various aspects to investigate its security risks. For example, De Groef et al. [38] study the identity authenticity of communicate peers and proposes several attack scenarios to compromise peers' identity authenticity, while [32] studies the IP leaking incurred by WebRTC deployment and profiles the extent to which the selection of browsers, VPN services, and VPN clients can affect this leaking issue. Moving forward, [42] works to prevent IP leaking through a browser extension and a network traffic proxy. Besides, other security risks have also been revealed, especially the risks to proxy peers' local network environment through in-browser network scanning [47], [66] and the potential abuse of peers' bandwidth resource [66]. Recently, Barradas in [36], explores new security applications of WebRTC and proposes Protozoa, a novel multimedia covert streaming channel, to circumvent network censorship leveraging in-browser WebRTC support. In our research, we reveal the wide use of WebRTC in peer-assisted video streaming services, and identify the peer IP leak risk.

**Resource squatting**. A line of works [37], [41], [62] have revealed cryptojacking wherein device computing resources are abused by miscreants for cryptocurrency mining. In addition, another abuse scenario is the unauthorized monetization of residential and mobile devices into web proxies to relay third-party network traffic [59], [60]. Moving forward from these studies, we reveal for the first time how video viewers' devices and network resources can be consumed without user consent to serve the video streaming services and third-party PDN providers.

## 8. Conclusions

In this paper, we carried out the first empirical study on the security risks of PDN ecosystem. Our study leads to the discovery of 3 representative PDN providers along with 134 websites, 38 mobile apps, and 9 private PDN services. Through a PDN analysis framework, we uncovered and evaluated four significant security risks, including service free riding risk, video segment pollution attack, peer IP leak risk, and resource squatting, which may affect millions of video viewers. Upon a solid understanding of these security risks, we have proposed several defense options to mitigate the risks.

## References

[1] Attack demo on content pollution. https://sites.google.com/view/pdnsec/home/demo.

[2] Chromedriver. https://chromedriver.chromium.org/.

[3] Develop with Docker Engine API. https://docs.docker.com/engine/api/.

[4] Geckodriver. https://github.com/mozilla/geckodriver.

[5] Http live streaming. https://tools.ietf.org/html/rfc8216.

[6] Jw player - hybrid peer-to-peer delivery. https://demos.jwplayer.com/peer-accelerated-delivery/.

[7] Live stream downloader. https://chrome.google.com/webstore/detail/live-stream-downloader/looepbdllpjgdmkpdcdffhdbmpbcfekj?hl=en.

[8] Peer5 docs - microsoft stream integration. https://docs.peer5.com/platforms/microsoft-stream/.

[9] Peer5 joins microsoft. https://blog.peer5.com/peer5-joins-microsoft/?ref=ms-banner.

[10] Peerjs - simple peer-to-peer with webrtc. https://peerjs.com/.

[11] Secure reliable transport. https://tools.ietf.org/html/draft-sharabayko-srt-00.

[12] Seleniumhq broswer automation. https://www.selenium.dev/.

[13] Tcpdump/libpcap public repository. https://www.tcpdump.org/.

[14] Webrtc. https://webrtc.org/.

[15] Webrtc turn server. https://webrtc.org/getting-started/turn-server.

[16] Spoify removes peer-to-peer technologyy from its deskop client. https://techcrunch.com/2014/04/17/spotify-removes-peer-to-peer-technology-from-its-desktop-client/, 2014.

[17] Cisco predicts more ip traffic in the next five years than in the history of the internet. Technical report, November 2018.

[18] Akama network deployment. https://www.akamai.com/us/en/about/facts-figures.jsp, 2021.

[19] Bogon filtering. https://en.wikipedia.org/wiki/Bogon_filtering, Apr. 2021.

[20] Content delivery nework:cdn mesh delivery—lumen. https://www.lumen.com/en-us/edge-computing/mesh-delivery.html?utm_source=Streamroot&utm_medium=link&utm_campaign=Streamroot-transition, 2021.

[21] Farsight security, cyber security intelligence solutions. https://www.farsightsecurity.com/, 2021.

[22] Ipinfo.io:comprehensive ip address data, ip geolocation api and database. https://ipinfo.io/, 2021.

[23] Nerdydata.com: Search the web's source code. https://www.nerdydata.com/, 2021.

[24] Peer5. https://www.peer5.com/product/, 2021.

[25] Publicwww.com: Search engine for sourcecode. https://publicwww.com/, 2021.

[26] Secure your streamroot key - streamroot documentation. https://support.streamroot.io/hc/en-us/articles/360021721214-Secure-your-Streamroot-Key, 2021.

[27] Security - Peer5 P2P Docs. https://docs.peer5.com/security/, 2021.

[28] Similarweb - similarweb traffic analysis. https://www.similarweb.com/, 2021.

[29] Viblast PDN. https://viblast.com/pdn/enterprise/, 2021.

[30] Web archive for peer5's homepage on march 5th, 2021. http://web.archive.org/web/20210305133908/https://www.peer5.com/, 2021.

[31] Youtube embedded players and player parameters. https://developers.google.com/youtube/player_parameters, Apr. 2021.

[32] Nasser Mohammed Al-Fannah. One leak will sink a ship: Webrtc ip address leaks. In *2017 International Carnahan Conference on Security Technology (ICCST)*, pages 1–5. IEEE, 2017.

[33] Kevin Allix, Tegawendé F Bissyandé, Jacques Klein, and Yves Le Traon. Androzoo: Collecting millions of android apps for the research community. In *2016 IEEE/ACM 13th Working Conference on Mining Software Repositories (MSR)*, pages 468–471. IEEE, 2016.

[34] Nasreen Anjum, Dmytro Karamshuk, Mohammad Shikh-Bahaei, and Nishanth Sastry. Survey on peer-assisted content delivery networks. *Computer Networks*, 116:79–95, 2017.

[35] François Baccelli, Fabien Mathieu, Ilkka Norros, and Rémi Varloot. Can p2p networks be super-scalable? In *2013 Proceedings IEEE INFOCOM*, pages 1753–1761. IEEE, 2013.

[36] Diogo Barradas, Nuno Santos, Luís Rodrigues, and Vítor Nunes. Poking a hole in the wall: Efficient censorship-resistant internet communications by parasitizing on webrtc. In *Proceedings of the 2020 ACM SIGSAC Conference on Computer and Communications Security*, pages 35–48, 2020.

[37] Hugo LJ Bijmans, Tim M Booij, and Christian Doerr. Inadvertently making cyber criminals rich: A comprehensive study of cryptojacking campaigns at internet scale. In *28th {USENIX} Security Symposium ({USENIX} Security 19)*, pages 1627–1644, 2019.

[38] Willem De Groef, Deepak Subramanian, Martin Johns, Frank Piessens, and Lieven Desmet. Ensuring endpoint authenticity in webrtc peer-to-peer communication. In *Proceedings of the 31st Annual ACM Symposium on Applied Computing*, pages 2103–2110, 2016.

[39] Prithula Dhungel, Xiaojun Hei, Keith W Ross, and Nitesh Saxena. The pollution attack in p2p live video streaming: measurement results and defenses. In *Proceedings of the 2007 workshop on Peer-to-peer streaming and IP-TV*, pages 323–328, 2007.

[40] Steven Englehardt and Arvind Narayanan. Online tracking: A 1-million-site measurement and analysis. In *Proceedings of the 2016 ACM SIGSAC conference on computer and communications security*, pages 1388–1401, 2016.

[41] Shayan Eskandari, Andreas Leoutsarakos, Troy Mursch, and Jeremy Clark. A first look at browser-based cryptojacking. In *2018 IEEE European Symposium on Security and Privacy Workshops (EuroS&PW)*, pages 58–66. IEEE, 2018.

[42] Alexandros Fakis, Georgios Karopoulos, and Georgios Kambourakis. Neither denied nor exposed: Fixing webrtc privacy leaks. *Future Internet*, 12(5):92, 2020.

[43] Ian Fette and Alexey Melnikov. The websocket protocol. RFC 6455, RFC Editor, December 2011.

[44] Gabriela Gheorghe, Renato Lo Cigno, and Alberto Montresor. Security and privacy issues in p2p streaming systems: A survey. *Peer-to-Peer Networking and Applications*, 4(2):75–91, 2011.

[45] Mikael Goldmann and Gunnar Kreitz. Measurements on the spotify peer-assisted music-on-demand streaming system. In *2011 IEEE International Conference on Peer-to-Peer Computing*, pages 206–211. IEEE, 2011.

[46] Dick Hardt. The oauth 2.0 authorization framework. Technical report, 2012.

[47] Mohammadreza Hazhirpasand and Mohammad Ghafari. One leak is enough to expose them all. In *International Symposium on Engineering Secure Software and Systems*, pages 61–76. Springer, 2018.

[48] Ethan Heilman, Alison Kendler, Aviv Zohar, and Sharon Goldberg. Eclipse attacks on bitcoin's peer-to-peer network. In *24th {USENIX} Security Symposium ({USENIX} Security 15)*, pages 129–144, 2015.

[49] Ping Identity and W Denniss. Oauth 2.0 token binding. 2017.

[50] Alan Johnston and Robert J. Sparks. Session description protocol (sdp) offer/answer examples. RFC 4317, RFC Editor, December 2005.

[51] Michael B Jones. The emerging json-based identity protocol suite. In *W3C workshop on identity in the browser*, pages 1–3, 2011.

[52] Ari Keranen, Christer Holmberg, and Jonathan Rosenberg. Interactive connectivity establishment (ice): A protocol for network address translator (nat) traversal. RFC 8445, RFC Editor, July 2018.

[53] Jian Liang, Rakesh Kumar, Yongjian Xi, and Keith W Ross. Pollution in p2p file sharing systems. In *Proceedings IEEE 24th Annual Joint Conference of the IEEE Computer and Communications Societies.*, volume 2, pages 1174–1185. IEEE, 2005.

[54] Jian Liang, Naoum Naoumov, and Keith W Ross. The index poisoning attack in p2p file sharing systems. In *INFOCOM*, pages 1–12. Citeseer, 2006.

[55] Eric Lin, Daniel Medeiros Nunes de Castro, Mea Wang, and John Aycock. Spoim: A close look at pollution attacks in p2p live streaming. In *2010 IEEE 18th International Workshop on Quality of Service (IWQoS)*, pages 1–9. IEEE, 2010.

[56] Ge Ma, Zhi Wang, Miao Zhang, Jiahui Ye, Minghua Chen, and Wenwu Zhu. Understanding performance of edge content caching for mobile video streaming. *IEEE Journal on Selected Areas in Communications*, 35(5):1076–1089, 2017.

[57] Ming Ma, Zhi Wang, Ke Su, and Lifeng Sun. Understanding content placement strategies in smartrouter-based peer video cdn. In *Proceedings of the 26th International Workshop on Network and Operating Systems Support for Digital Audio and Video*, pages 1–6, 2016.

[58] Yuval Marcus, Ethan Heilman, and Sharon Goldberg. Low-resource eclipse attacks on ethereum's peer-to-peer network. *IACR Cryptol. ePrint Arch.*, 2018:236, 2018.

[59] Xianghang Mi, Xuan Feng, Xiaojing Liao, Baojun Liu, XiaoFeng Wang, Feng Qian, Zhou Li, Sumayah Alrwais, Limin Sun, and Ying Liu. Resident evil: Understanding residential ip proxy as a dark service. In *2019 IEEE Symposium on Security and Privacy (SP)*, pages 1185–1201. IEEE, 2019.

[60] Xianghang Mi, Siyuan Tang, Zhengyi Li, Xiaojing Liao, Feng Qian, and XiaoFeng Wang. Your phone is my proxy: Detecting and understanding mobile proxy networks. 2021.

[61] Guillaume Montassier, Thibault Cholez, Guillaume Doyen, Rida Khatoun, Isabelle Chrisment, and Olivier Festor. Content pollution quantification in large p2p networks: A measurement study on kad. In *2011 IEEE International Conference on Peer-to-Peer Computing*, pages 30–33. IEEE, 2011.

[62] Marius Musch, Christian Wressnegger, Martin Johns, and Konrad Rieck. Thieves in the browser: Web-based cryptojacking in the wild. In *Proceedings of the 14th International Conference on Availability, Reliability and Security*, pages 1–10, 2019.

[63] H Parmar and M Thornburgh. Adobe's real time messaging protocol. *Copyright Adobe Systems Incorporated*, pages 1–52, 2012.

[64] Victor Le Pochat, Tom Van Goethem, Samaneh Tajalizadehkhoob, Maciej Korczyński, and Wouter Joosen. Tranco: A research-oriented top sites ranking hardened against manipulation. In *proceedings of the Network and Distributed System Security Symposium*, 2019.

[65] A Rahmatulloh, R Gunawan, and FMS Nursuwars. Performance comparison of signed algorithms on json web token. In *IOP Conference Series: Materials Science and Engineering*, volume 550, page 012023. IOP Publishing, 2019.

[66] Andreas Reiter and Alexander Marsalek. Webrtc: your privacy is at risk. In *Proceedings of the Symposium on Applied Computing*, pages 664–669, 2017.

[67] Roberto Roverso and Mikael Högqvist. Hive. js: Browser-based distributed caching for adaptive video streaming. In *2014 IEEE International Symposium on Multimedia*, pages 143–146. IEEE, 2014.

[68] Henning Schulzrinne, Anup Rao, and Robert Lanphier. Real time streaming protocol (rtsp), 1998.

[69] Prajakta Solapurkar. Building secure healthcare services using oauth 2.0 and json web token in iot cloud scenario. In *2016 2nd International Conference on Contemporary Computing and Informatics (IC3I)*, pages 99–104. IEEE, 2016.

[70] Thomas Stockhammer. Dynamic adaptive streaming over http– standards and design principles. In *Proceedings of the second annual ACM conference on Multimedia systems*, pages 133–144, 2011.

[71] Haizhou Wang, Xingshu Chen, Wenxian Wang, and Mei Ya Chan. Content pollution propagation in the overlay network of peer-to-peer live streaming systems: modelling and analysis. *IET Communications*, 12(17):2119–2131, 2018.

[72] Jason Weil, Victor Kuarsingh, Chris Donley, Christopher Liljenstolpe, and Marla Azinger. Iana-reserved ipv4 prefix for shared address space. *IETF Request for Comment*, 6598, 2012.

[73] Hao Yin, Xuening Liu, Tongyu Zhan, Vyas Sekar, Feng Qiu, Chuang Lin, Hui Zhang, and Bo Li. Livesky: Enhancing cdn with p2p. *ACM Transactions on Multimedia Computing, Communications, and Applications (TOMM)*, 6(3):1–19, 2010.

[74] Wei Yu, Corey Boyer, Sriram Chellappan, and Dong Xuan. Peer-to-peer system-based active worm attacks: Modeling and analysis. In *IEEE International Conference on Communications, 2005. ICC 2005. 2005*, volume 1, pages 295–300. IEEE, 2005.

[75] Ge Zhang, Wei Liu, Xiaojun Hei, and Wenqing Cheng. Unreeling xunlei kankan: Understanding hybrid cdn-p2p video-on-demand streaming. *IEEE Transactions on Multimedia*, 17(2):229–242, 2014.

[76] Liang Zhang, Fangfei Zhou, Alan Mislove, and Ravi Sundaram. Maygh: Building a cdn from client web browsers. In *Proceedings of the 8th ACM European Conference on Computer Systems*, pages 281–294, 2013.

[77] Mingchen Zhao, Paarijaat Aditya, Ang Chen, Yin Lin, Andreas Haeberlen, Peter Druschel, Bruce Maggs, Bill Wishon, and Miroslav Ponec. Peer-assisted content distribution in akamai netsession. In *Proceedings of the 2013 conference on Internet measurement conference*, pages 31–42, 2013.

[78] Roverli P Ziwich, Elias P Duarte Jr, and Glaucio P Silveira. Distributed mitigation of content pollution in peer-to-peer video streaming networks. *IET Communications*, 2020.

## 9. Appendix

### 9.1. Identified PDN Signatures & Customers

Table 5 presents the identified signatures of PDN providers. Table 6, Table 7, and Table 8 list respectively top public PDN websites, top public PDN apps, and 9 private PDN services, as identified in our study.

TABLE 5: Signatures of the studied PDN providers

| PDN Provider | Web Signatures | APK Signatures |
|---|---|---|
| Peer5 | api.peer5.com/peer5.js?id=* <br> api.peer5.com/peer5*.plugin.js | com.peer5.android <br> com.peer5.ApiKey <br> com.peer5.sdk |
| Streamroot | cdn.streamroot.io/*.js?srKey=* <br> cdn.streamroot.io/*dna-wrapper.js | io.streamroot.dna <br> io.streamroot.dna.StreamrootKey |
| Viblast | viblast.js <br> data-viblast-key <br> data-viblast-enable-pdn | com.viblast.android <br> libviblast.so <br> libnative-viblast.jni.so |
| Private | RTCPeerConnection & <br> RTCIceCandidate & <br> iceServers | – |

TABLE 6: Top 15 detected PDN websites

| PDN websites | PDN provider | # Monthly visits |
|---|---|---|
| dailymail.co.uk | Peer5 | 373M |
| rt.com | Streamroot | 117M |
| gamespot.com | Peer5 | 78M |
| clarin.com | Peer5 | 69M |
| terra.com.br | Peer5 | 55M |
| cnet.com | Peer5 | 46M |
| rtve.es | Peer5 | 35M |
| tn.com.ar | Peer5 | 21M |
| tvguide.com | Peer5 | 18M |
| reallifecam.com | Viblast | 17M |
| manyvids.com | Streamroot | 17M |
| metacritic.com | Peer5 | 16M |
| wetter.de | Streamroot | 15M |
| n1info.com | Viblast | 14M |
| voyeur-house.tv | Viblast | 14M |

TABLE 7: Top 15 detected PDN apps

| PDN apps | PDN provider | # Google Play downloads |
|---|---|---|
| com.graymatrix.did | Peer5 | 100M |
| iflix.play | Streamroot | 50M |
| com.fplay.activity | Peer5 | 10M |
| com.nousguide.android.rbtv | Peer5 | 10M |
| com.portonics.mygp | Peer5 | 10M |
| mivo.tv | Peer5 | 10M |
| vn.vtv.vtvgo | Peer5 | 10M |
| com.tdcm.trueidapp | Peer5 | 10M |
| com.tru | Peer5 | 10M |
| vn.vtv.vtvgotv | Peer5 | 5M |
| com.bongo.bioscope | Peer5 | 5M |
| tv.fubo.mobile | Peer5 | 5M |
| tv.sweet.tvplayer | Streamroot | 1M |
| com.tvplayer | Streamroot | 1M |
| com.ug.eon.android | Viblast | 1M |

TABLE 8: 9 confirmed private PDN services

| PDN websites | Country | # Monthly visits |
|---|---|---|
| ok.ru | Russia | 662M |
| douyu.com | China | 95M |
| v.qq.com | China | 92M |
| iqiyi.com | China | 82M |
| huya.com | China | 61M |
| youku.com | China | 60M |
| tudou.com | China | 44M |
| mgtv.com | China | 42M |
| younow.com | US | 1M |

## 9.2. Generating Polluted Video Segments

To circumvent the constraints in the *video segment pollution test* mentioned in §4.3, we propose a general method to generate polluted video segments: overlay the target video on top of the original video files. Specifically, the attacker first retrieves all video segment files and combines them as the original video file. Then the attacker overlays the target text or video on top of the original video file to obtain the polluted file. Since the target text or video is on the top level, it will be displayed to viewers. Then the attacker can split the polluted video into segments with the same parameters as the original video file. The splitting parameters, i.e., the length of each segment and the segment filename, can be easily retrieved from the original manifest file. As a result, the attacker can generate the polluted video segments and play them seamlessly with the original manifest file.

## 9.3. Evaluation of Peer-assisted Integrity Checking

We set up a simulation environment to evaluate our mitigation against the video segment pollution in terms of feasibility and performance. Our simulation environment includes a signaling server, a PDN JavaScript SDK, and a website integrating this SDK. Among these components, the signaling server and the PDN SDK are built upon an open-source WebRTC library, PeerJS [10], since there is no open-sourced PDN system available up to our knowledge. Leveraging this simulation environment, we implemented our defense designs and demonstrated their feasibility. We further evaluated the performance overhead with a focus on profiling the resource consumption of IM calculation and verification. This is achieved through three groups of control experiments. In each group, we specified 6 peers, with 3 as the senders and the other 3 as the receivers. Each receiver peer requests from the senders a typical video segment with a length of 10 seconds, lasting for a total of 600 seconds. Different groups are set based on settings including whether to do P2P video segment delivery, and whether to do IM calculation for the sending peers and IM verification for the receiving peers. During this process, the resource consumption of the sending peers is measured using Docker APIs, while the latency of IM calculation is measured by the time difference of $T_{recv} - T_{send}$, where $T_{recv}$ is the receiving time after IM verification and $T_{send}$ means the sending time before IM calculation. The results are shown in Table 9. As we can see, the IM calculation will incur negligible CPU and memory consumption on average. And the latency of adding IM checking is increased by less than 80 ms for a video segment of 3MB size.

TABLE 9: Evaluation for IM checking

| Browser | PDN | IM checking | CPU | Memory | Latency |
|---|---|---|---|---|---|
| Chrome | No | No | 1 | 1 | - |
| Chrome | Yes | No | 1.11 | 1.21 | 67ms |
| Chrome | Yes | Yes | 1.14 | 1.24 | 140ms |